\documentclass[acmtog]{acmart}
\usepackage{booktabs} % For formal tables
\usepackage{enumitem}         % Customisation des itemize et enumerate

\setcitestyle{square}

\usepackage[ruled]{algorithm2e} % For algorithms

\SetAlFnt{\small}
\SetAlCapFnt{\small}
\SetAlCapNameFnt{\small}
\SetAlCapHSkip{0pt}
\IncMargin{-\parindent}

\setcopyright{acmlicensed}
\acmJournal{TOG}
\acmYear{2018}
\acmVolume{37}
\acmNumber{6}
\acmArticle{266}
\acmMonth{11}
\acmDOI{10.1145/3272127.3275037}

\title{There are 174 Subdivisions of the Hexahedron into Tetrahedra}

\author{Jeanne Pellerin}
\orcid{0000-0001-8481-7509}
\affiliation{%
  \institution{Universit\'e catholique de Louvain}
  \streetaddress{Avenue Georges Lema\^itre 4-6}
  \city{Louvain-la-Neuve}
  \postcode{1348}
  \country{Belgique}
}%
 \email{jeanne.pellerin@uclouvain.be}

\author{Kilian Verhetsel}
\affiliation{%
  \institution{Universit\'e catholique de Louvain}
  \streetaddress{Avenue Georges Lema\^itre 4-6}
  \city{Louvain-la-Neuve}
  \postcode{1348}
  \country{Belgique}
 }%
 \email{kilian.verhetsel@uclouvain.be}

\author{Jean-Fran\c cois Remacle}
\orcid{0000-0002-4798-6458}
\affiliation{%
  \institution{Universit\'e catholique de Louvain}
  \streetaddress{Avenue Georges Lema\^itre 4-6}
  \city{Louvain-la-Neuve}
  \postcode{1348}
  \country{Belgique}
 }%
 \email{jean-francois.remacle@uclouvain.be}

\begin{abstract}
This article answers an important theoretical question: How many different
subdivisions of the hexahedron into tetrahedra are there? It is well known that
the cube has five subdivisions into 6 tetrahedra and one subdivision into 5 tetrahedra.
However, all hexahedra are not cubes and moving the vertex positions increases the number
of subdivisions. Recent hexahedral dominant meshing methods try to take these
configurations into account for combining tetrahedra into hexahedra, but fail to enumerate
them all: they use only a set of 10 subdivisions among the 174 we found in this article.

The enumeration of these 174 subdivisions of the hexahedron into tetrahedra is our combinatorial result.
Each of the 174 subdivisions has between 5~and~15 tetrahedra and is actually a class
of 2 to 48 equivalent instances which are identical up to vertex relabeling.
We further show that exactly 171 of these subdivisions
have a geometrical realization, i.e. there exist coordinates of the eight
hexahedron vertices in a three-dimensional space such that the geometrical tetrahedral mesh is valid.
We exhibit the tetrahedral meshes for these configurations
and show in particular subdivisions of hexahedra
with 15 tetrahedra that have a strictly positive Jacobian.

\end{abstract}



\begin{document}

\maketitle

\section{Introduction}
 
Engineering analysis is generally based on finite element computations and requires a mesh.
Compared to tetrahedral meshes, hexahedral meshes have important numerical properties: faster
assembly \citep{remacle_gpu_2016}, high accuracy in solid mechanics \citep{wang_back_2004},
or for quasi-incompressible materials \citep{benzley_comparison_1995}.
Unstructured hexahedral meshing is an active research topic, e.g.~\citep{gao_robust_2017, lyon_hexex:_2016, solomon_boundary_2017,fang_all-hex_2016}.
However, no robust hexahedral meshing technique is yet able to process general 3D domains,
and the generation of unstructured hexahedral finite element meshes remains
the biggest challenge in mesh generation \citep{shepherd_hexahedral_2008}.

Several promising methods propose to leverage the existence of robust and efficient
tetrahedral meshing algorithms e.g.~\citep{si_tetgen_2015}
to generate hexahedral meshes by combining tetrahedra into hexahedra
\citep{meshkat_generating_2000, yamakawa_fully-automated_2003, levy_l_2010,
huang_boundary_2011, baudouin_frontal_2014, botella_indirect_2016, sokolov_hexahedral-dominant_2016, pellerin_identifying_2018}.
It was recently suggested by \citep{pellerin_identifying_2018} that the
set of 10 subdivisions into tetrahedra of the hexahedron used by most of these methods is incomplete.

\begin{figure}[h]
	\centering
	\includegraphics[width=.70\linewidth]{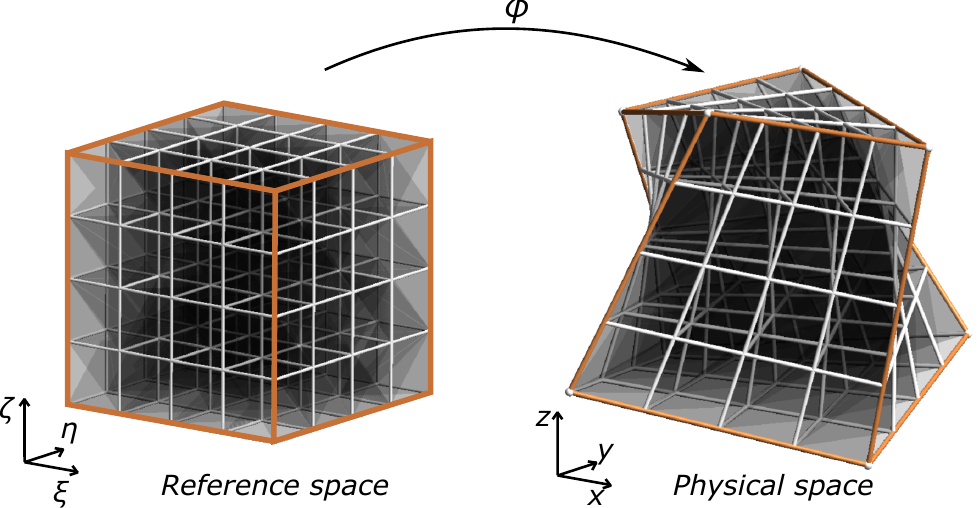}
	\caption{The hexahedra we study (right) are the image of a reference cube (left) by a trilinear mapping.}
	\label{fig:trilinear_hex}
\end{figure}

In this paper, a hexahedron is combinatorially a 3-dimensional cube
whose square facets have been triangulated and whose geometry is defined as
 a finite hexahedral element is the image of a reference cube
by a trilinear mapping denoted $\varphi$ on Figure~\ref{fig:trilinear_hex}.
Our hexahedra are not strictly cubes, moreover they are only valid as
long as the mapping from the reference cube is injective.
Validity is difficult to evaluate because
the Jacobian determinant is not constant over the hexahedron e.g.~\citep{chandrupatla_introduction_2011}.

In this work, we consider the elementary subdivisions of the hexahedron in
tetrahedra. In the remaining of this paper, we use the generic term of
triangulation to refer to $d$-dimensional simplicial subdivisions, i.e. tetrahedral or triangular meshes (triangulations).
The faces of the hexahedron are subdivided into two triangles. As we will see in Section~\ref{sec:related_work},
no complete enumeration of the hexahedron subdivisions into tetrahedra is available, except for
the specific case of the cube \citep{de_loera_triangulations:_2010}.

Our first contribution is to solve the purely  combinatorial problem of the hexahedron triangulations.
Our result is that there are 174 triangulations of the hexahedron up to isomorphism
that have between 5 and 15 tetrahedra (Table~\ref{tab:nb_triangulations}).
Depending on symmetries, each of the 174 triangulations is actually a class
of 2, 4, 8, 12, 16, 24, or 48 equivalent instances, which are identical up to vertex relabeling.
We prove this result by reviewing all the
combinatorial triangulations of the 3-ball with eight vertices.

 \begin{table}
	\setlength{\tabcolsep}{3.5pt}
	\centering
	\caption{
			Numbers of combinatorial triangulations of the hexahedron with 5 to 15 tetrahedra up to isomorphism.
	}
	\label{tab:nb_triangulations}
	\begin{tabular}{lcccccccccccc}
		\toprule
		\#tets & 5 & 6 & 7 & 8 & 9 & 10 & 11& 12 & 13 & 14 & 15   & Sum \\
		\midrule
		combinatorial & 1 & 5 & 5 & 7 & 13 & 20 & 35 & 30 & 28 & 19 & 11 & 174 \\
		geometrical & 1 & 5 & 5 & 7 & 13 & 20 & 35 & 30 & 28 & 19 & 8 & 171 \\
		\bottomrule
		\end{tabular}
\end{table}

Our second contribution is to demonstrate that exactly 171 hexahedron triangulations have a geometrical realization in $\mathbb{R}^3$.
To solve the realization problem we implemented the method proposed in \citep{firsching2017realizability}.
In particular, we exhibit the meshes of hexahedra with 14 and 15 tetrahedra contradicting the
belief that there might only be subdivisions with up to 13 tetrahedra \citep{meshkat_generating_2000, sokolov_hexahedral-dominant_2016}.
This is an important theoretical result that links hexahedral meshes to tetrahedral meshes.
We attached our implementation and all the result files of combinatorial triangulations
and tetrahedral meshes in the supplementary materials.

%-----------------------------------------------------------------------------------------------------------------------------------
%-----------------------------------------------------------------------------------------------------------------------------------
\section{Related work}
\label{sec:related_work}
%-----------------------------------------------------------------------------------------------------------------------------------
%-----------------------------------------------------------------------------------------------------------------------------------

Previous works on the subdivisions into tetrahedra of the hexahedron are limited to the specific configuration of the cube.
They were studied in discrete geometry \citep{de_loera_triangulations:_2010} and in mesh generation methods
 that combine tetrahedra into hexahedra \citep{meshkat_generating_2000, botella_indirect_2016, sokolov_hexahedral-dominant_2016}.

\begin{figure}[h]
	\centering
	\includegraphics[width=.95\linewidth]{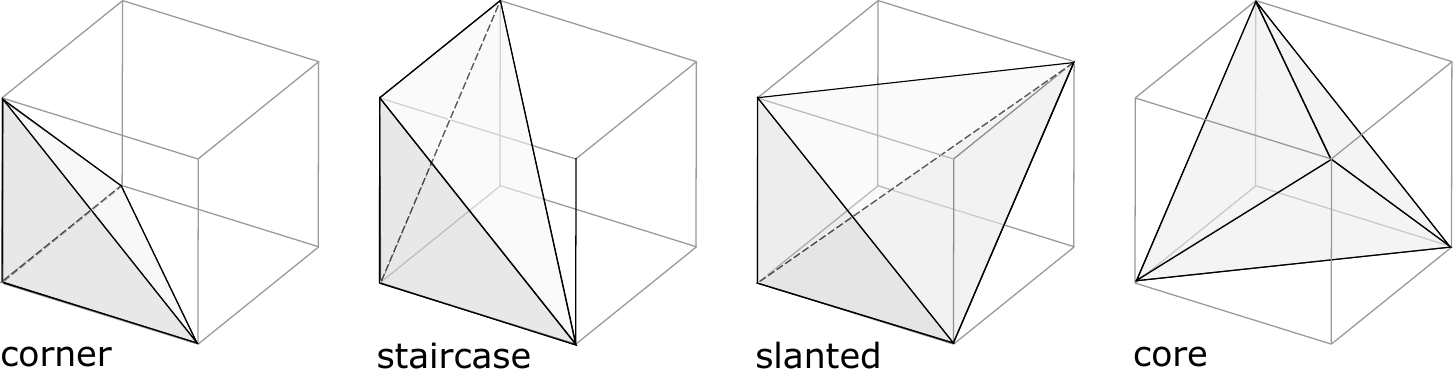}
	\caption{The four types of tetrahedra in a cube $I^3 = [0,1]^3$.}
	\label{fig:cube_tet_types}
\end{figure}
\paragraph{The cube}
Following \citep{de_loera_triangulations:_2010}, let us first look at the
different types of tetrahedra that can be built from the eight vertices of a cube  $I^3 = [0,1]^3$.
Suppose that one facet of the tetrahedron is contained in one of the six cube facets,
there are four possible vertices to built a non-degenerate tetrahedra that represent three cases
due to symmetry (Figure~\ref{fig:cube_tet_types}):

\begin{itemize}
	\item \emph{Corner} tetrahedra have three facets on the cube faces.
	\item \emph{Staircase} tetrahedra have two facets on the cube faces.
	\item \emph{Slanted} tetrahedra have a unique facet on the cube faces.
\end{itemize}
The last type of tetrahedra are \emph{Core} tetrahedra that have no facet on the cube faces.
For the 3-cube, this list is complete and irredundant \citep{de_loera_triangulations:_2010}.

\begin{figure}
	\centering
	\includegraphics[width=0.8\linewidth]{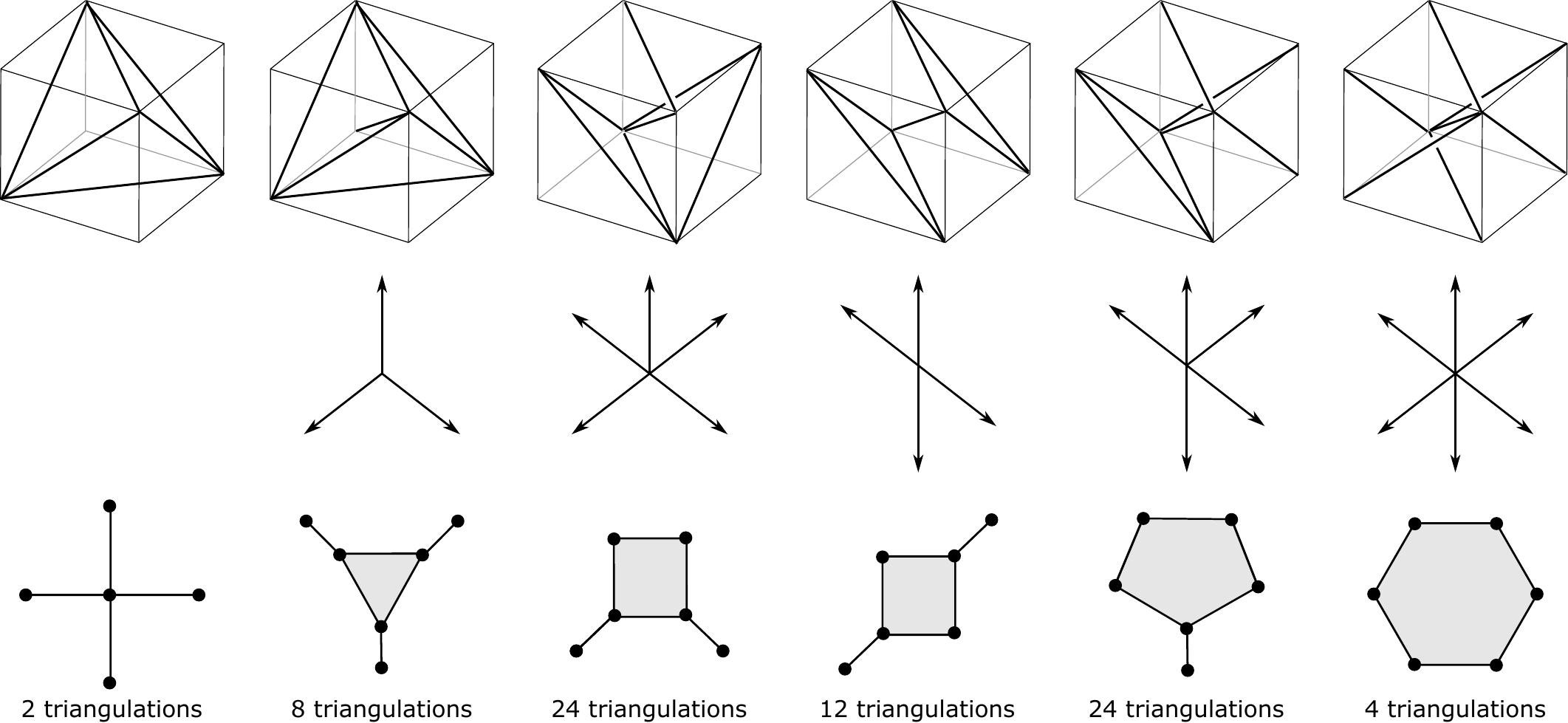}
	\caption{The six triangulations of the 3-cube $I^3 = [0,1]^3$ up to isomorphism and their dual complexes.
	Each class corresponds to 2 to 24 equivalent triangulations.}
	\label{fig:cube_triangulations}
\end{figure}

From this classification of the tetrahedra in a cube triangulation follows the classification of triangulations of the 3-cube.
The 3-cube has exactly 74 triangulations that are classified in 6 classes (Figure~\ref{fig:cube_triangulations}).
\begin{itemize}
	\item Every triangulation of the 3-cube contains either a regular tetrahedron
	(i.e. a tetrahedron whose 6 edges are of equal lengths) or a diameter, i.e. an interior edge joining two
	opposite vertices.

	\item There are two triangulations of the first type, symmetric to one another.
	The triangulations of the second type are completely classified,
	modulo symmetries by their dual complexes (Figure~\ref{fig:cube_triangulations}).
\end{itemize}

The dual complex of a triangulation is a graph in which there is a vertex for each tetrahedron
and one edge if corresponding tetrahedra share a triangle (Figure~\ref{fig:cube_triangulations}).
By construction, a 2-cell of the dual complex corresponds to an interior edge of the triangulation.

There are at least two ways to prove that these six triangulations are
indeed the six possible types of triangulations of the cube.
Discrete geometers rely on the determination of the cell containing the barycenter
of the cube and on symmetry considerations about the vertex arrangement around that cell \citep{de_loera_triangulations:_2010}.
\citep{meshkat_generating_2000} enumerate the possible dual complexes.

\paragraph{The 3-cube in finite precision}
Floating point numbers have a finite precision
and represent the cube $I^3 = [0,1]^3$ up to some tolerance.
Moreover, sets of points that are exactly cospherical and coplanar are often specifically processed
because these non-general positions are problematic for many geometrical algorithms
 \citep{edelsbrunner_simulation_1990}.
Therefore, to combine elements of tetrahedral meshes,
\citep{botella_indirect_2016} and \citep{sokolov_hexahedral-dominant_2016}
consider subdivisions of the cube into two prisms linked by a \textit{sliver}, i.e. a very flat tetrahedra
connecting four points that are coplanar up to a given tolerance.
Note that \citep{sokolov_hexahedral-dominant_2016} aim at enumerating the triangulations of more general configurations,
but the geometrical argument used in the proof limits its validity
to the cube represented in finite precision.

\begin{figure}[h]
	\centering
	\includegraphics[width=0.5\linewidth]{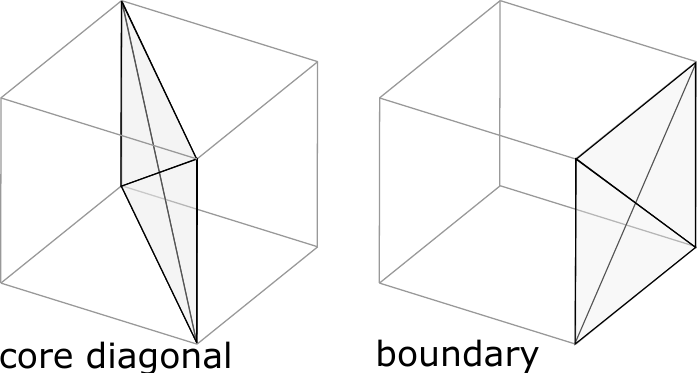}
	\caption{Two types of tetrahedra in a finite precision cube.
	}
	\label{fig:cube_7_tet_types}
\end{figure}

\begin{figure}[h]
	\centering
	\includegraphics[width=.95\linewidth]{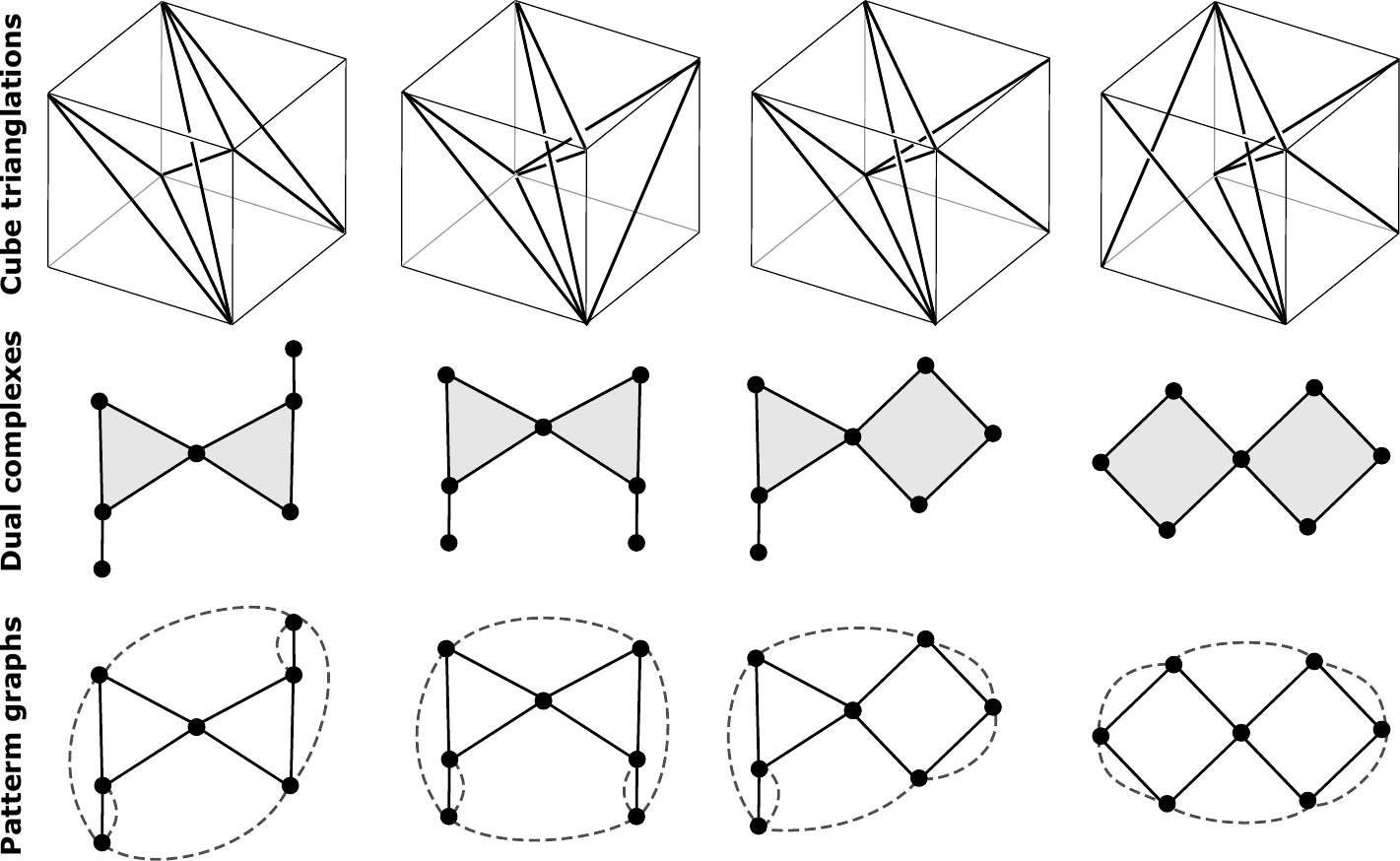}
	\caption{
		The four triangulations of the finite precision cube with seven tetrahedra.
		Dashed edges connect pairs of tetrahedra incident to the same hexahedron facet.
	}
	\label{fig:cube_7_tet_triangulations}
\end{figure}

In a finite precision 3-cube, two new types of non-degenerate tetrahedra appear (Figure~\ref{fig:cube_7_tet_types}):
\begin{itemize}
	\item \emph{Core diagonal} tetrahedra have no facet on the hexahedron
	boundary. They subdivide the hexahedron into two prisms,
	creating four additional types of triangulations into seven tetrahedra (Figure~\ref{fig:cube_7_tet_triangulations}).

	\item \emph{Boundary} tetrahedra connect the four vertices of a facet of the cube. 
	 We exclude these tetrahedra from the triangulations we consider
	 because they are not strictly inside the hexahedron.
	 Indeed, the hexahedron facets are bilinear patches strictly inside boundary tetrahedra.
	
\end{itemize}
Together with the previous four types of tetrahedra in the cube 
they constitute the only possible configurations for a tetrahedron in a hexahedron triangulation.

%-----------------------------------------------------------------------------------------------------------------------------------
%-----------------------------------------------------------------------------------------------------------------------------------
\section{Combinatorial Triangulations} 
\label{sec:patterns}
%-----------------------------------------------------------------------------------------------------------------------------------
%-----------------------------------------------------------------------------------------------------------------------------------

\begin{figure}[h]
	\centering
	\includegraphics[width=.3\linewidth]{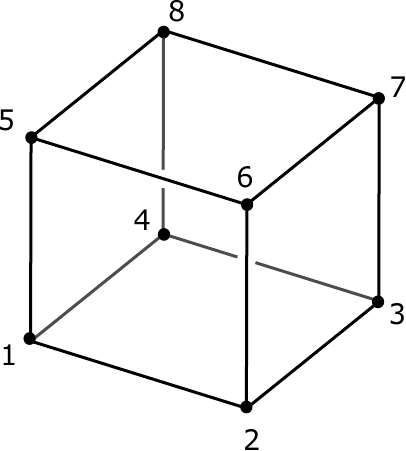}
	\caption{Hexahedron template}
	\label{fig:hex_connectivity}
\end{figure}

The geometrical problem of the triangulation of the hexahedron
is  difficult and we first tackle the purely combinatorial side of the question.
A set of tetrahedra is a valid combinatorial triangulation of the hexahedron $\{12345678\}$ 
(Figure~\ref{fig:hex_connectivity})
if it is a valid combinatorial 3-manifold with 8 vertices and if its boundary
contains the 8 vertices, 12 edges \{12\}, \{14\}, \{15\}, \{23\}, \{26\}, \{34\}, \{36\}, \{48\}, \{56\}, \{58\}, \{67\}, \{78\}
and 12 triangular facets that can be paired into 6 quadrilaterals \{1234\}, \{1265\}, \{1485\}, \{2376\}, \{3487\}, \{5678\}.

\begin{theorem}
\label{th:mytheorem}
	The hexahedron $\{12345678\}$ has 174 combinatorial triangulations up to isomorphism  that
	do not contain any boundary tetrahedra.
\end{theorem}
The 174 triangulations are enumerated in Table~\ref{tab:all_triangulations} and we provide
their complete description in  supplemental material.

To demonstrate Theorem~\ref{th:mytheorem} we
first use a result about the triangulations of the 3-sphere with 9 vertices
to compute all triangulations of the 3-ball with 8~vertices (Section~\ref{sec:3sphere}).
At the next step, we select triangulations whose boundary is the boundary of a triangulated hexahedron (Section~\ref{sec:hex_boundary}).
Finally the valid combinations are classified into isomorphism classes (Section~\ref{sec:decomposition_graph}).

\begin{table*}
 \scriptsize
 \caption{
	The 174 combinatorial triangulations of the hexahedron per number of tetrahedra and corresponding
	hexahedron surface triangulations (Figure~\ref{fig:cube_7_tet_triangulations}).
	3~combinatorial triangulations do not have a geometrical realization (bold).
	13~do not have a convex geometrical realization in $\mathbb{R}^3$ (underlined).
	All combinatorial and geometrical triangulations are provided in the supplemental material.
  }
 \label{tab:all_triangulations}
   \begin{tabular}{*{22}{l}}
   \toprule
  &      &       &       &               &      &               &       &               &        &                &      & 11\_AI &  g &                  &          &               &       &                &    &       &       \\
  &      &       &       &               &      &               &       &               &        &                &      & 11\_AH &  f &                  &          &               &       &                &    &       &       \\
  &      &       &       &               &      &               &       &               &        &                &      & 11\_AG &  g &                  &          &               &       &                &    &       &       \\
  &      &       &       &               &      &               &       &               &        &                &      & 11\_AF &  d &                  &          &               &       &                &    &       &       \\
  &      &       &       &               &      &               &       &               &        &                &      & 11\_AE &  g &                  &          &               &       &                &    &       &       \\
  &      &       &       &               &      &               &       &               &        &                &      & 11\_AD & g &12\_AD &  c &               &       &                &    &       &       \\
  &      &       &       &               &      &               &       &               &        &                &      & 11\_AC &  g &12\_AC &  d &               &       &                &    &       &       \\
  &      &       &       &               &      &               &       &               &        &                &      & 11\_AB &  b &12\_AB &  g & 13\_AB& g &                &    &       &       \\
  &      &       &       &               &      &               &       &               &        &                &      & 11\_AA &  d &12\_AA &  d & 13\_AA& g &                &    &       &       \\
  &      &       &       &               &      &               &       &               &        &                &      & 11\_Z  &  g  &12\_Z  &  g & 13\_Z & g &                &    &       &       \\
  &      &       &       &               &      &               &       &               &        &                &      & 11\_Y  &  g  &\underline{12\_Y}  &  g & 13\_Y & g &                &    &       &       \\
  &      &       &       &               &      &               &       &               &        &                &      & 11\_X  &  f &12\_X  &  d & 13\_X & g &                &    &       &       \\
  &      &       &       &               &      &               &       &               &        &                &      & 11\_W  & g &12\_W  &  g & \underline{13\_W} & f &                &    &       &       \\
  &      &       &       &               &      &               &       &               &        &                &      & 11\_V  &  g &12\_V  &  d & 13\_V & d &                &    &       &       \\
  &      &       &       &               &      &               &       &               &        &                &      & 11\_U  &  g &\underline{12\_U}  &  d & 13\_U & c &                &    &       &       \\
  &      &       &       &               &      &               &       &               &        & 10\_T &    f& 11\_T  &  g &12\_T   & g& \underline{13\_T} & c &                &    &       &       \\
  &      &       &       &               &      &               &       &               &        & 10\_S &    d& 11\_S  &  d &12\_S  & g& 13\_S & d & 14\_S & g &       &       \\
  &      &       &       &               &      &               &       &               &        & 10\_R &    g & 11\_R  &  e &12\_R  &g& 13\_R & d & \underline{14\_R} & g &       &       \\
  &      &       &       &               &      &               &       &               &        & 10\_Q &   g & 11\_Q  &  g &12\_Q  &g & 13\_Q & d & 14\_Q & g &       &       \\
  &      &       &       &               &      &               &       &               &        & 10\_P &    g & 11\_P  &  e &12\_P  &  f & 13\_P & d & \underline{14\_P} & d &       &       \\
  &      &       &       &               &      &               &       &               &        & 10\_O &    f& 11\_O  &  d &12\_O  &  d & 13\_O & g & \underline{14\_O} & c &       &       \\
  &      &       &       &               &      &               &       &               &        & 10\_N &    f& 11\_N  &  e &12\_N  &  g & 13\_N & g & \underline{14\_N} & d &       &       \\
  &      &       &       &               &      &               &       &  9\_M&   f   & 10\_M &    g& 11\_M  &  e &12\_M  &  g & 13\_M & d & 14\_M & c &       &       \\
  &      &       &       &               &      &               &       &  9\_L&   f   & 10\_L &    b& 11\_L  &  g &12\_L  &  e & 13\_L & g & 14\_L & c &       &       \\
  &      &       &       &               &      &               &       &  9\_K&   f   & 10\_K &    f& 11\_K  &  g &12\_K  &  e & 13\_K & d & 14\_K & d & 15\_K &    g \\
  &      &       &       &               &      &               &       &  9\_J&   d   & 10\_J &    g& 11\_J  &  f&12\_J  &  g & 13\_J & e & 14\_J & d & \textbf{15\_J} &    g \\
  &      &       &       &               &      &               &       &  9\_I&   g   & 10\_I &    g& 11\_I  & f  &12\_I  &  d & 13\_I & e & 14\_I & d & \underline{15\_I} &    d \\
  &      &       &       &               &      &               &       &  9\_H&   d   & 10\_H &    d& 11\_H  &  g &12\_H  &  e & 13\_H & d & 14\_H & f & \underline{15\_H} &    c \\
  &      &       &       &               &      &  8\_G&   f  &  9\_G&   b   & 10\_G &    g& 11\_G  &  g &12\_G  &  e & 13\_G & e & 14\_G & d & \underline{15\_G} &    c \\
  &      &               &       &               &      &  8\_F&   b  &  9\_F&   f   & 10\_F &    d& 11\_F  &  g &12\_F  &  e & 13\_F & d & 14\_F & e & \underline{15\_F} &    c \\
  &   & 6\_E& a &  7\_E&   c &  8\_E&   d  &  9\_E&   d   & 10\_E &    g& 11\_E  &  c &12\_E  &  g & 13\_E & f & 14\_E & e & 15\_E &   c \\
  &   & 6\_D&c  &  7\_D&  b &  8\_D&   d  &  9\_D&  g  & 10\_D &    d& 11\_D  &  g &12\_D  &  g & 13\_D & d & 14\_D & e & 15\_D &  f \\
  &  &  6\_C&   b  &  7\_C&   g &  8\_C&   g  &  9\_C&   d   & 10\_C &    e& 11\_C  &  d&12\_C  &  g & 13\_C & g & 14\_C & d & 15\_C & d \\
  &  & 6\_B&    e &  7\_B&   d &  8\_B&   g  &  9\_B&   e   & 10\_B &    e& 11\_B  &  f &12\_B  &  g & 13\_B & g & 14\_B & d & \textbf{15\_B} & f \\
  5\_A&  a   &  6\_A&   d  &  7\_A&   b &  8\_A&   d  &  9\_A&   d   & 10\_A &  g& 11\_A  &  g &12\_A  &  d & 13\_A & f & \underline{14\_A} & c & \textbf{15\_A} &    g \\

  \bottomrule
  \end{tabular}%
\end{table*}

%-----------------------------------------------------------------------------------------------------------------------------------
\subsection{Triangulations of the 3-Ball}
\label{sec:3sphere}
%-----------------------------------------------------------------------------------------------------------------------------------

\begin{figure}[h]
	\centering
	\includegraphics[width=.8\linewidth]{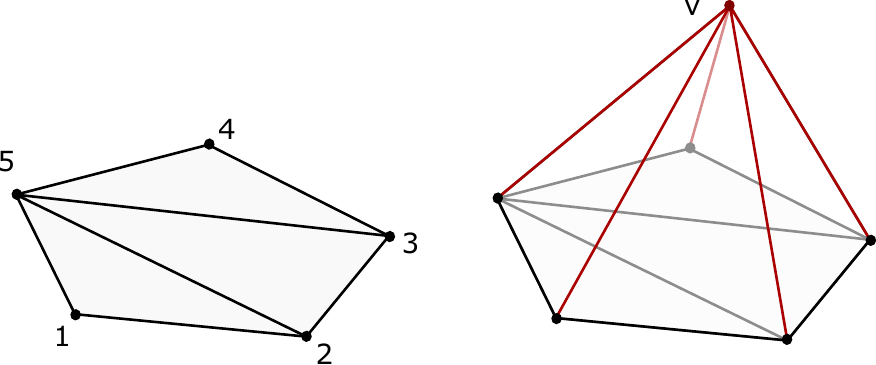}
	\caption{
		Building a cone over a triangulation of a 2-dimensional pentagon into 3 triangles gives a triangulation of the 2-sphere
		into 8 triangles.
	}
	\label{fig:cone}
\end{figure}

The sphere, or 2-sphere, is the 2-dimensional surface boundary  of a 3-dimensional ball.
The 3-sphere generalizes this, and  is defined as the 3-dimensional boundary of a 4-dimensional ball.
There is a direct link between the triangulations of the 2-sphere (3-sphere) and the
triangulations of the 2-ball  (3-sphere). Indeed, a triangulation of the 2-sphere can be
constructed from the triangulation of a 2-ball by building a cone as illustrated on Figure~\ref{fig:cone}.
The cone is built by adding point $v$ over a polygon and connecting it to the five boundary
edges.
The inverse transformation, the removal of one point $v$ of the sphere triangulation
as well as all triangles incident to $v$, permits to obtain the triangulation of a ball.
(The interested reader is referred to \citep{ziegler_lectures_1995} for more details on this classical construction).

In this paper, we enumerate all triangulations of the hexahedron, i.e. 3-dimensional ball with 8 vertices.
They can be obtained from the triangulations of the 3-sphere with 9 vertices.
The 1296 triangulations of the 3-sphere have been enumerated by \citep{altshuler_neighborly_1973, altshuler_neighborly_1974,altshuler_enumeration_1976,altshuler_classification_1980}
and are available online \citep{lutz_manifold_2018}.
Nine triangulations of the 3-ball with eight vertices
can be built from each of the 1296 triangulations by
removing one of the vertices $v_i, i={1,\dots,9}$ and its link, i.e. all tetrahedra incident to $v_i$.

%-----------------------------------------------------------------------------------------------------------------------------------
\subsection{Triangulations of the Hexahedron}
\label{sec:hex_boundary}
%-----------------------------------------------------------------------------------------------------------------------------------

\begin{figure}[h]
	\centering
	\includegraphics[width=\linewidth]{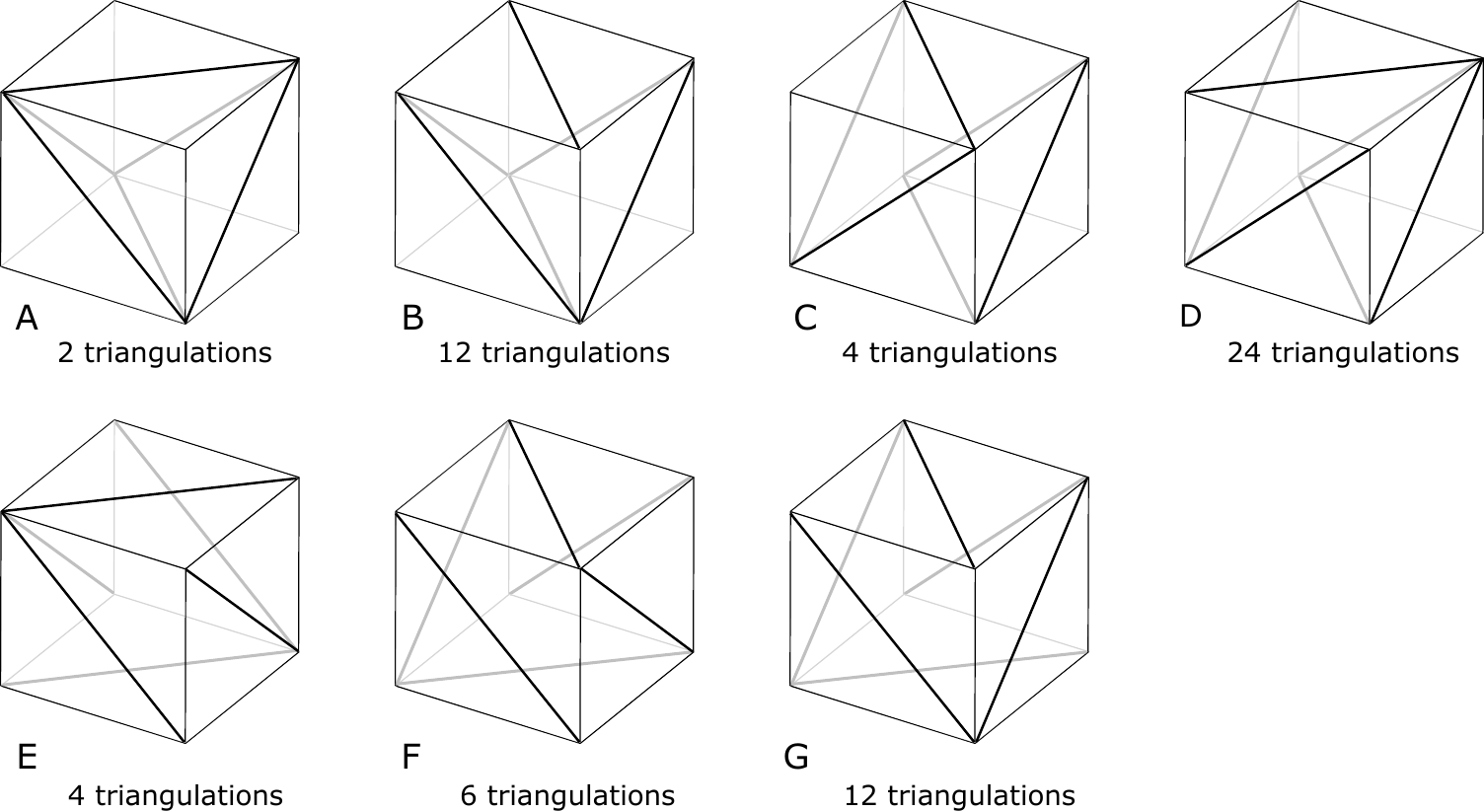}
	\caption{The 7 triangulations of a hexahedron boundary up to isomorphism.
   }
	\label{fig:hex_boundary_7_triangulations}
\end{figure}

The next step is to test if the obtained 3-ball triangulations are valid hexahedron triangulations.
They are if their boundary matches the triangulated boundary of a hexahedron.
The triangulation of the boundary of a hexahedron has 8 vertices
and 18 edges. Among these, 12 are fixed and there are 2 possibilities
to place the remaining 6 diagonals of the quadrilateral facets.
We have then $2^6 = 64$ possible triangulations.
These triangulations can be classified into 7 equivalence classes,
i.e. there are 7 triangulations of the hexahedron boundary up to isomorphism (Figure~\ref{fig:hex_boundary_7_triangulations}).

%
%-----------------------------------------------------------------------------------------------------------------------------------
\subsection{Hexahedron Triangulation Comparison}
\label{sec:decomposition_graph}
%-----------------------------------------------------------------------------------------------------------------------------------
The final step is to compare the obtained triangulations of the hexahedron
and determine those that are the same up to isomorphism.
We use a  graph formalism proposed in \citep{meshkat_generating_2000}
to represent and compare the hexahedron triangulations.

 \begin{figure}[h]
	\centering
	\includegraphics[width=0.8\linewidth]{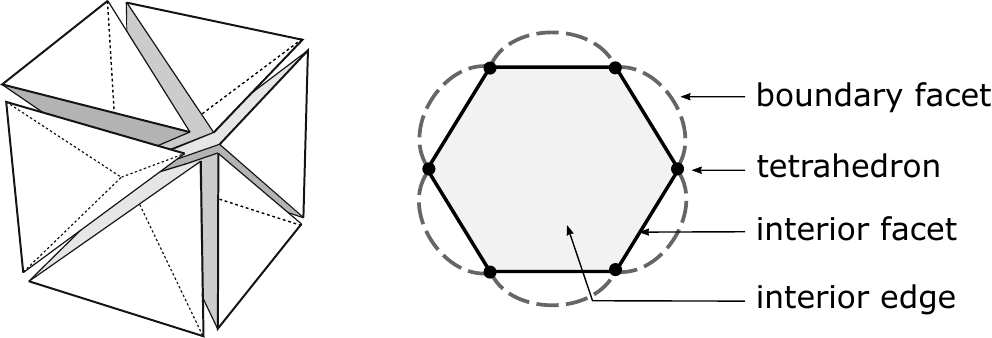}
	\caption{Definition of the decomposition graph that is used to represent hexahedron triangulations. }
	\label{fig:decomposition_pattern_definition}
\end{figure}
 The \emph{decomposition graph} (Figure~\ref{fig:decomposition_pattern_definition})
of the triangulation of a hexahedron is an edge-colored graph where there is one vertex per tetrahedron,
one black (plain) edge between vertices if the corresponding tetrahedra are adjacent,
and one grey (dashed) edge between vertices if the corresponding tetrahedra have triangle faces on the same hexahedron facet.
By construction, each vertex is of degree 4 and simple chordless cycles of plain edges correspond to interior edges of the triangulation.

\begin{figure}
	\centering
	\includegraphics[width=.9\linewidth]{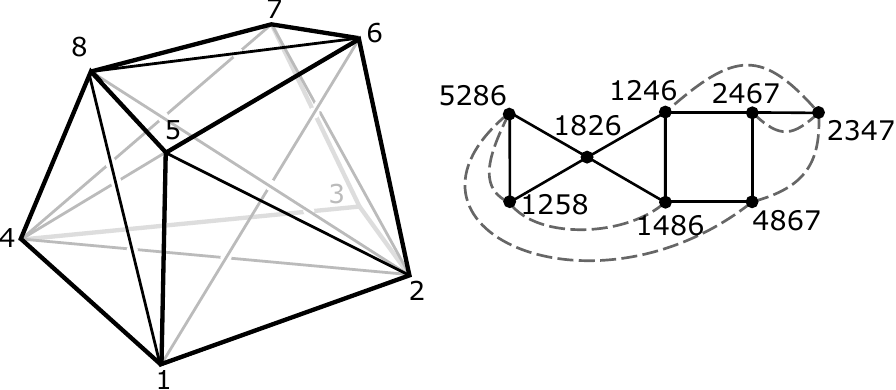}
	\caption{The correspondence between a hexahedron triangulations and its decomposition graph for triangulation 8\_E.
	}
	\label{fig:8_E}
\end{figure}

There is a one-to-one correspondence between triangulations without boundary tetrahedra and decomposition graphs (Figure~\ref{fig:8_E}).
Suppose that we fix the vertex labels of the triangulation such that \{12345678\} is a hexahedron.
Then 12 of its edges are fixed (\{12\}, \{14\}, \{15\}, \{23\}, \{26\}, \{34\}, \{36\}, \{48\}, \{56\}, \{58\}, \{67\}, \{78\}).
There is then one choice to subdivide each of the 6 facets into two triangles.
Once the 6 boundary diagonal edges are chosen, the remaining degrees of freedom to triangulate
the hexahedron are controlled by the interior facets.
The plain edges of the decomposition graph represent the interior facets, while the 6 dashed edges
represent the diagonal edges.

Two combinatorial triangulations are isomorphic, if and only if their decomposition graphs are isomorphic,
 i.e. are the same with respect to the relabeling of their vertices.
More formally, an isomorphism of graphs $G$ and $H$ is a bijection between
the vertex sets of $G$ and $H$, $f:V(G) \rightarrow V(H)$, such that any two vertices
$u$ and $v$ of $G$ are adjacent if and only if $f(u)$ and $f(v)$ are adjacent in~$H$.

The decomposition graphs of the triangulations of the hexahedron obtained
in the previous section are therefore classified into isomorphism classes
to obtain our main combinatorial result.
Our reference implementation, attached in the supplementary materials,
uses the nauty library \cite{mckay_practical_2014}.
This concludes the demonstration of Theorem~\ref{th:mytheorem}.

%-----------------------------------------------------------------------------------------------------------------------------------
%-----------------------------------------------------------------------------------------------------------------------------------
\section{Geometrical triangulations}
\label{sec:geometrical_realizations}
%-----------------------------------------------------------------------------------------------------------------------------------
%-----------------------------------------------------------------------------------------------------------------------------------

We now tackle the geometrical side of the problem of the triangulations of the hexahedron in $\mathbb{R}^3$.
There is indeed no guarantee that a valid geometrical triangulation exists for each of the 174 combinatorial triangulations.

More formally, we are solving a geometrical realization problem for each combinatorial triangulation $T$,
i.e. we are searching positions of the 8 vertices, $\mathcal{A}$
in $\mathbb{R}^{3}$, such that the geometrical triangulation defined by $T$ is valid.
A triangulation is valid if  (1) the union of the tetrahedra is the convex hull of the points
and (2) there is no unwanted intersection between tetrahedra.

\begin{theorem}
\label{th:mysecondtheorem}
	171 of the 174 combinatorial triangulations of the hexahedron have a geometrical
	realization in $\mathbb{R}^3$.
\end{theorem}
The 171 triangulations are enumerated in Table~\ref{tab:all_triangulations}
and we provide the realizations in the supplemental material.
The 3 triangulations that do not have a realization have 15 tetrahedra and are illustrated on Figure~\ref{fig:all_15_configs}.

\begin{algorithm}

	\KwData{$T$ combinatorial triangulation}
	\KwResult{$\mathcal{A}$ a realization}
	\BlankLine
	clauses = encodeSAT( $T$ )  \tcc*{Section 4.2} \
	$S$ = solveSAT(clauses) \tcc*{MiniSAT solver} \
	\If{ $ S==\emptyset$}{
		\tcc{No oriented matroid, not realizable}
		\Return $\emptyset$\;
	}
	 \ForEach{$\chi$ in  S}{
		inequalities = setDeterminantInequalities($\chi$) \tcc*{Section 4.3}
		$\mathcal{A}$ = solveSystem(inequalities)  \tcc*{SCIP solver}
		\If{$\mathcal{A} \neq \emptyset$}{
			\tcc{Realization is found}
			\Return $\mathcal{A}$ \;
		}
    }
	\tcc{Check non realizability}
	undecidedRealizability = false\;
	\ForEach{$\chi \in$  S}{
		\lIf{ ! \rm{findFinalPolynomial}($\chi$) \tcc*{Section 4.4} }{undecidedRealizability = true}
	}
	\caption{Geometrical realization}
	\label{algo:realization}
\end{algorithm}

Our proof follows the strategy described in \citep{joswig_computing_2003, schewe2010nonrealizable, firsching2017realizability}
which consists in (1) solving the underlying combinatorial problem of the point configurations
that can realize a given triangulation $T$, before (2) finding point coordinates in $\mathbb{R}^3$ corresponding to one configuration.
The main steps are described in Algorithm~\ref{algo:realization}.

%-----------------------------------------------------------------------------------------------------------------------------------
\subsection{Definitions}
\label{sec:matroid}
%-----------------------------------------------------------------------------------------------------------------------------------

To encode the combinatorial properties of the point sets realizing a given combinatorial
triangulation, we use oriented matroids. This is an important topic of discrete mathematics \citep{bjorner1999oriented},
here we restrict definitions to the studied problem.

Let $E$ be a finite set, $r \in \mathbb{N}$ with $ r \leq |E| $, and
a mapping $\chi : E^r \mapsto \{-1, +1\}$.
$\mathcal{M} = (E, \chi)$ is a \emph{uniform oriented matroid} of rank $r$
if the following two conditions are satisfied:

\begin{enumerate}
\item The mapping $\chi$ is alternating, i.e. for all permutations $\pi$ of $\{1, \dots, r\}$
  \[\chi(e_{\pi(1)}, e_{\pi(2)}, \dots,e_{\pi(r)}) =
    sgn(\pi)\chi(e_{1},e_{2}, \dots, e_ {r})\]

where $sgn(\pi)$ is the permutation sign, i.e. 1 for even permutations
and -1 for odd permutations.

\item For all $\sigma \in {{E}\choose{r-2}}$ and all subsets
$\{e_1, e_2, e_3, e_4\}  \subseteq E \setminus \sigma$, we have that:

\[\begin{split}
  \{-1, +1\} \subseteq \{
    \chi(\sigma, e_1, e_2)\chi(\sigma, e_3, e_4), \\
    -\chi(\sigma, e_1, e_3)\chi(\sigma, e_2, e_4), \\
    \chi(\sigma, e_1, e_4)\chi(\sigma, e_2, e_3)
    \}
\end{split}\]

\end{enumerate}

The mapping $\chi$ is called the \emph{chirotope} of the oriented matroid.
Given a set of points $\mathcal{A}$ in $\mathbb{R}^d$, we can build a chirotope
for a matroid of rank $r = d+1$ by considering the determinant of the points in homogeneous coordinates
using $v \mapsto {{1}\choose{v}}$.

Set for any $(\mathbf{a_1}, \mathbf{a_2}, \dots, \mathbf{a_{d+1}}) \in \mathcal{A}$
\[\chi\left(1,2, \dots, d+1\right) = {\sf sign}
  \det \begin{pmatrix} 1 & 1 & \dots & 1 \\
    \mathbf{a_1} & \mathbf{a_2} & \dots & \mathbf{a_{d+1}}
  \end{pmatrix}\]

\begin{figure}[h]
	\center
	\includegraphics[width =0.9\linewidth]{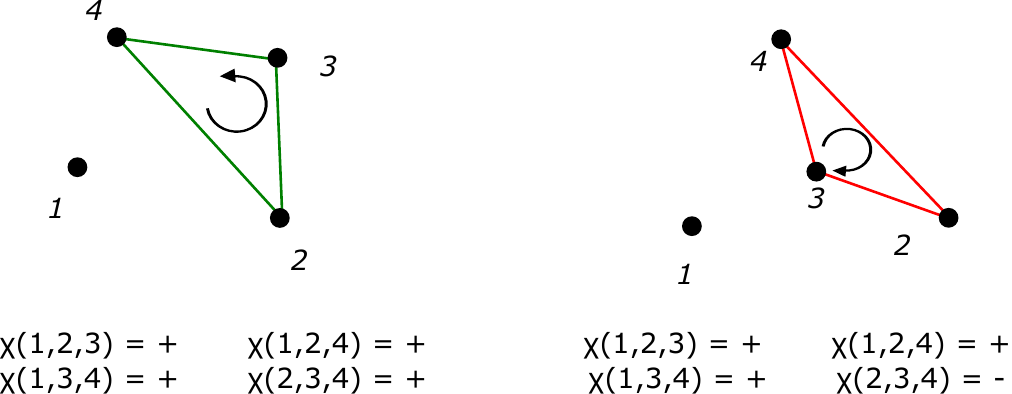}
	\caption{
		The  two possible 4-point configurations in $\mathbb{R}^2$ have different chirotopes.
		The chirotope encodes the orientation of all 4 triangles, only the one whose orientation changes is drawn.
	}
	\label{fig:chirotope}
\end{figure}

The chirotope is a sign function which gives the orientation of
all possible subsets of $r = d+1$  of points in $\mathcal{A}$.
The chirotope of a point set $\mathcal{A} \in \mathbb{R}^{2}$ (respectively $\mathbb{R}^{3}$) is
the orientation of all the triangles (respectively tetrahedra) that can be built from $\mathcal{A}$ (Figure \ref{fig:chirotope}).
The chirotope encodes the point set structure, and, as we will see in Section~\ref{sec:chirotope},
all necessary information to check if $\mathcal{A}$ is a realization of the combinatorial triangulation $T$.

Note that we only consider point sets $\mathcal{A}$ in general position (no 4-coplanar of 5-copsherical points) 
restricting ourselves to uniform oriented matroids, $\chi$ cannot take the value zero.

\subsection{Combinatorial Formulation}
\label{sec:chirotope}

We first review the five constraints that the oriented matroids of
point sets realizing a combinatorial triangulation $T$ must enforce, before
seeing how the oriented matroids are determined.

\paragraph{Constraints}

The point set is in general position therefore
the corresponding oriented matroid should :
\begin{enumerate}
	\item be uniform, $\chi \neq 0$  (no 4-coplanar or 5-cospherical points)
	\item be acyclic, a property shared by the oriented matroids of all point sets.
	\item be valid, all  $r-2$ subsets should satisfy condition (2) of the definition in Section~\ref{sec:matroid}.
\end{enumerate}

The tetrahedra of the geometrical triangulation are valid, therefore the oriented matroid should:
\begin{enumerate} \setcounter{enumi}{3}
	\item define valid tetrahedra, $\forall \sigma \in T, \chi(\sigma_{1}, \dots, \sigma_{d+1}) = +1$.
	We assume that they are consistently oriented and are all positive in $T$.
\end{enumerate}

The geometrical triangulation is valid, i.e. all tetrahedra of $T$ intersect properly, i.e.
for all pairs of tetrahedra  $\sigma$ and $\sigma'$ of $T$ we have $conv \, \sigma \cap conv \, \sigma' = conv (\sigma \cap \sigma')$.
The intersection property can also be translated to constraints on the oriented matroids, let us see how.

If a point set $\mathcal{A}$ is in general position, any subset of $d+1$ points
is affinely independent while any subset of $d+2$  points is affinely dependent,
i.e. there exists $\lambda \in \mathbb{R}^{d+2}$ such that:
\[\sum_{i = 1}^{d+2} \lambda_i a_i = \mathbf{0} \mbox{ and } \sum_{i = 1}^ {d + 2} \lambda_i = 0\]
We can then build two subsets of $\mathcal{A}$, $A^+ = \{a_i \;|\; \lambda_i > 0\}$ and $A^- = \{p_i \;|\; \lambda_i < 0\}$,
whose convex hulls intersect at a unique  point (Radon's theorem).
If two different simplices $\sigma$ and $\sigma'$ of $T$ intersect, then there is
a subset of $d+2$ points of $vert(\sigma) \cup vert(\sigma')$
that are affinely  dependent and corresponds to a pair $(X^+, X^-)$.
In oriented matroid jargon, this pair $(X^+, X^-)$ is called a circuit of $\mathcal{A}$,
a fundamental notion related to another definition of oriented matroids \citep{bjorner1999oriented}.
All ${{n}\choose{d+2}}$ subsets of $(d + 2)$ points of $\mathcal{A}$ define a circuit  $(X^+, X^-)$ that
can be computed from
\begin{align*}
X^+ &= \{i \;|\; (-1)^i\chi(1, 2, \dots, i-1,
i+1, \dots, d+2) = +1 \} \\
X^- &= \{i \;|\; (-1)^i\chi(1, 2, \dots, i-1,
i+1, \dots, d+2) = -1 \}
\end{align*}

This give us the final constraint on oriented matroids of points sets realizing a combinatorial triangulation $T$, they should:
\begin{enumerate} \setcounter{enumi}{4}
	\item not admit any circuit $(X^+, X^-)$ such that
           	$X^+ \subseteq \sigma$ and $X^- \subseteq \sigma'$ where $\sigma$ and $\sigma'$ are two tetrahedra of $T$.
\end{enumerate}

\paragraph{Solving}

The five constraints described above fix partially the values
taken by $\chi$ for all ${n \choose d + 1}$ subsets of point labels.
To completely determine the admissible oriented matroids for a triangulation $T$,
we follow \cite{schewe2010nonrealizable}, and solve an instance of a boolean satisfiability problem SAT
that is satisfiable, if and only if, an admissible oriented matroid exists for a triangulation $T$.
The details on the encoding of the constraints are given in  \cite{schewe2010nonrealizable}.
We used the MiniSAT \cite{sorensson2005minisat} solver to find all admissible
chirotopes for 174 combinatorial triangulations of the hexahedron.
Our implementation is provided in the supplemental material.

\paragraph{Result} Three combinatorial triangulations $15\_A, 15\_B, 15\_J$ do not admit
an oriented matroid that enforce the five constraints and are therefore not realizable.

\begin{figure*}
	\centering
	\includegraphics[width=0.9\textwidth]{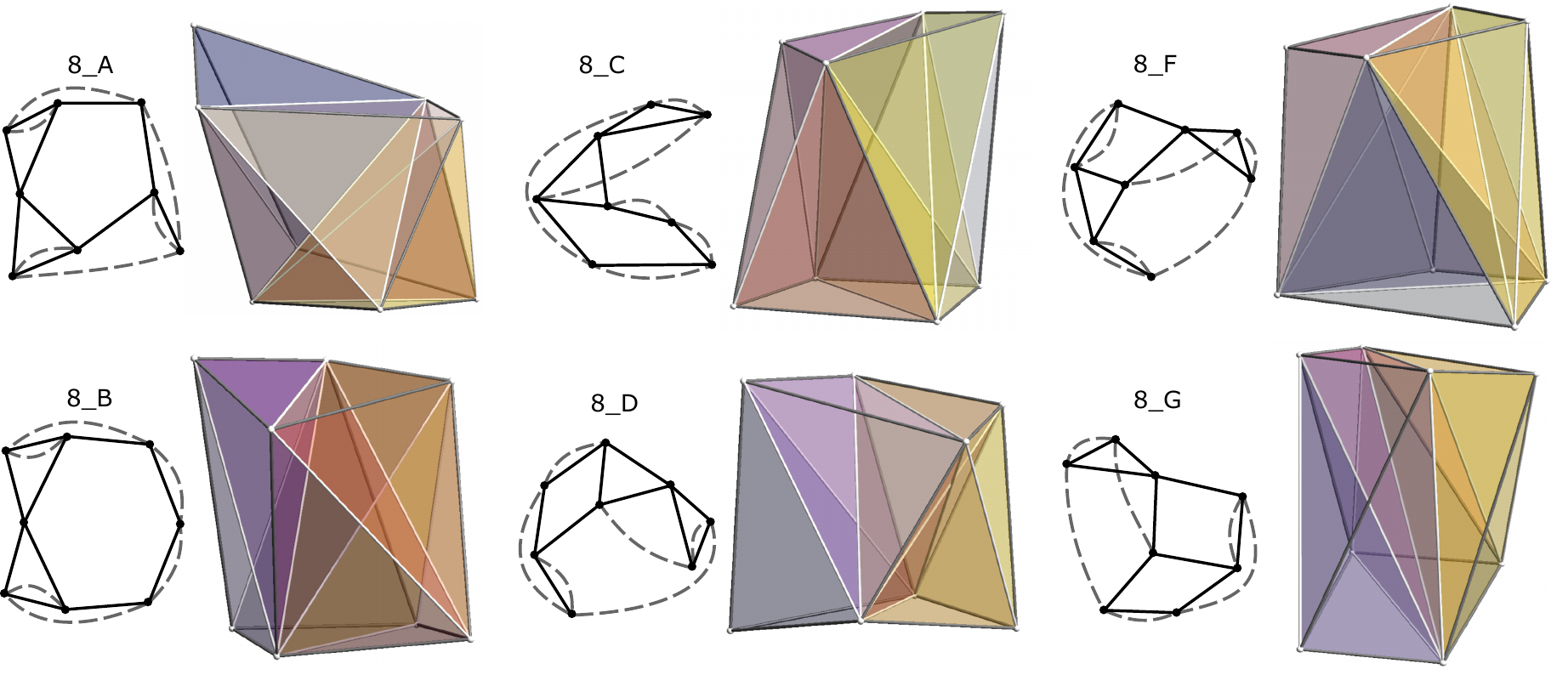}
	\caption{
		Six triangulations of the hexahedon into 8 tetrahedra and their geometrical realizations.
		Colors correspond to tetrahedron indices. Hexahedron edges are grey, while tetrahedron edges are white.
		The seventh triangulation with 8 tetrahedra, $8\_E$, is presented on Figure~\ref{fig:8_E}.
	}
	\label{fig:all_8_configs}
\end{figure*}

\subsection{Geometric Realizations}
\label{sec:scip}

Once the oriented matroids fulfilling all the constraints related to a given combinatorial triangulation are computed,
the final step is to find real coordinates for the 8 vertices realizing one of these
 oriented matroids.
We follow the approach of \citep{firsching2017realizability}
who proposes to solve the system of ${n \choose d+1}$
polynomial inequalities of the form:

\[\chi(1, 2, \dots, d+1) \det \begin{pmatrix} 1 & 1 & \dots & 1 \\
    \mathbf{p}_1 & \mathbf{p}_2 & \dots & \mathbf{p}_{d+1} \end{pmatrix} > 0 \]

where the possible values of $\chi(1, 2, \dots, d+1)$ have been computed in Section~\ref{sec:chirotope}.

\paragraph{Solving}
In order to find feasible solutions we used SCIP,  a solver that
relies on branch and bound techniques \cite{scip2017}.
To encode the strict inequalities numerically, we rewrite the
strictly positive constraint as superior or equal to an $\epsilon$ value
(we used $\epsilon = 10^{-4}$).
When a solution exists, SCIP quickly converges and we obtain a realization of the
combinatorial triangulation $T$.

\paragraph{Result}
All the 171 triangulations that admit an oriented matroid  have a realization.
All realizations are provided along our implementation in supplemental material.
This completes the demonstration of Theorem~\ref{th:mysecondtheorem}.
Figure~\ref{fig:all_8_configs} gives realizations for the configurations with 8 tetrahedra.

\subsection{Convex Geometric Realization}
We furthermore studied the realization problem with an additional
constraint on the 8 point positions to ensure that they are in
convex position, i.e. all vertices are on their convex hull  (none is in its interior).
This is a reasonable condition for hexahedra to be
valid for finite element computations.
The convexity condition can be directly taken
into account when testing admissible chirotopes by excluding all configurations in which
a vertex is inside a tetrahedron defined by four other vertices.

\begin{figure}
	\centering
	\includegraphics[width=0.71\linewidth]{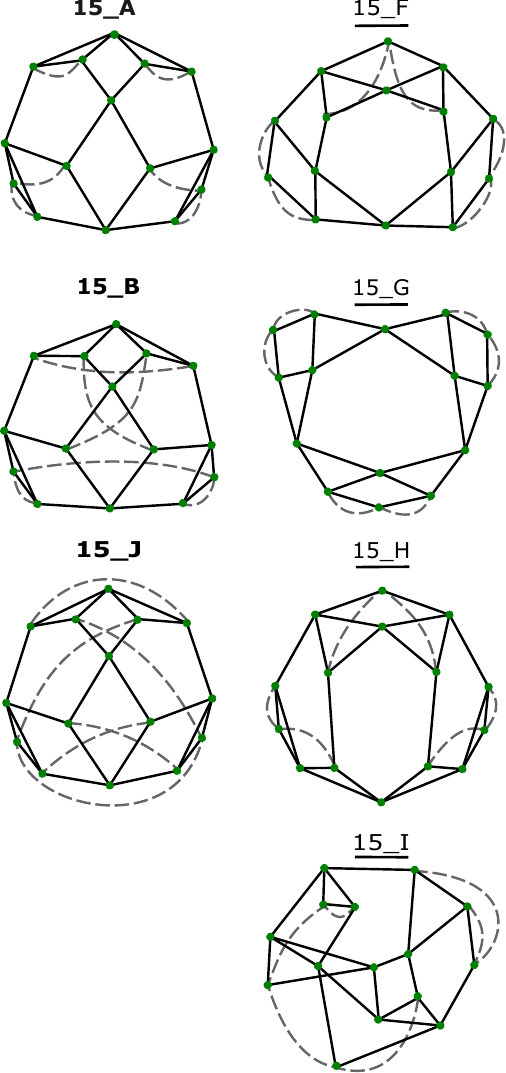}
	\caption{
		Three of the 11 combinatorial triangulations of the hexahedron into 15 tetrahedra
		are not realizable (left).
		Four combinatorial triangulations are realizable but with points in non-convex position (right).
	}
	\label{fig:all_15_configs}
\end{figure}

\paragraph{Result} Over the 171 realizable triangulations of the hexahedron,
13 do not have a realization such that their vertices are in convex position.
$12\_U, 12\_Y,$
$13\_T, 13\_W, $
$14\_A, 14\_N, $
$14\_O, 14\_P,$
$14\_R, 15\_F,$
$15\_H, 15\_I$
do not admit a uniform oriented matroid and are therefore not realizable.
Triangulation $15\_G$ admits an oriented matroid, but SCIP does not terminate
when searching for a realization (Figure~\ref{fig:all_15_configs}).
To prove that there is none, a different approach is necessary.
A property of non-realizable matroids is
that they admit a final polynomial, which can be found, if bi-quadratic,
using the practical method of \cite{bokowski1990finding}.
It consists in proving the infeasibility of a linear program
that encodes the second property of oriented matroid definition.
$15\_G$ oriented matroid has a final polynomial and is therefore
not realizable with points in convex position.

\begin{figure}
	\centering
	\includegraphics[width=0.7\linewidth]{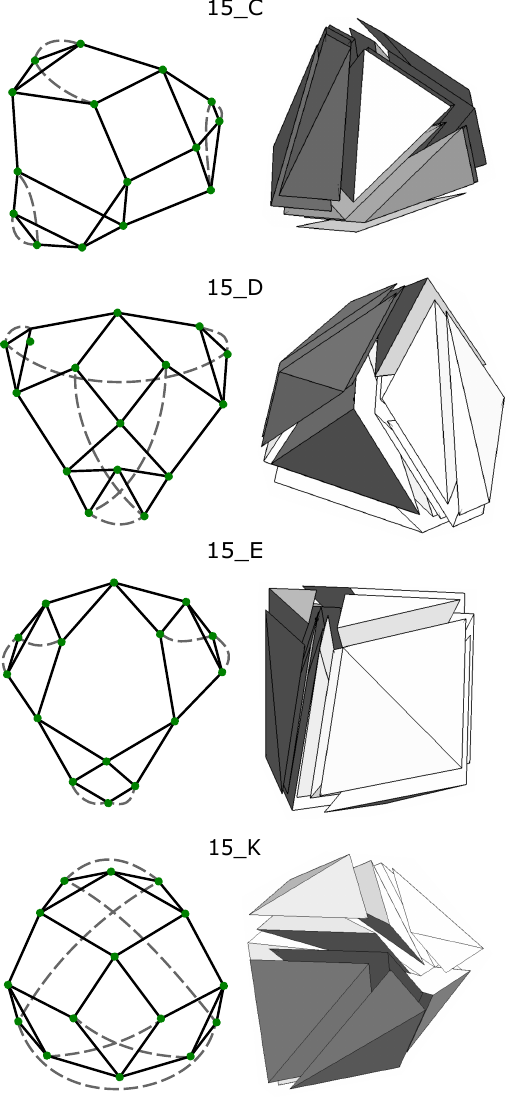}
	\caption{
		The four realizable triangulations with 15 tetrahedra and with points in convex position.
		The hexahedra are valid, their Jacobian is strictly positive.
	}
	\label{fig:realizable_15_configs}
\end{figure}

\section{Discussion}

We demonstrated  that there are 174 combinatorial triangulations
of the hexahedron up to isomorphism and that 171 of them have a geometric realization.

Our results are consistent with the previous results on the triangulations of the 3-cube
presented in \citep{de_loera_triangulations:_2010}.
However, Theorem 4 in \citep{sokolov_hexahedral-dominant_2016}, that states that
there are only 10 triangulations of the hexahedron with 5, 6, or 7 tetrahedra, is incorrect.
We described 161 additional geometrical triangulations of the hexahedron.
The smallest triangulations for which no valid hexahedron may exist have 12 tetrahedra.
The largest triangulations for which a valid hexahedron exists have 15 tetrahedra, by valid we mean a
a strictly positive Jacobian (Figure~\ref{fig:realizable_15_configs}).
The existence of these triangulations contradicts the belief that there
may only be subdivisions of the hexahedron with up to 13 tetrahedra.
The practical consequence is that the methods solely based on the 10 triangulations
of the cube to identify groups of tetrahedra that can be merged into a hexahedron
do not compute all the possible hexahedra as claimed
\citep{meshkat_generating_2000, yamakawa_fully-automated_2003, botella_indirect_2016, sokolov_hexahedral-dominant_2016}.
Adapting these methods to take into account dozens more patterns is theoretically possible,
but consequences on performances would be catastrophic.
Note that the alternative algorithm of \citep{pellerin_identifying_2018}
is not impacted by this observation, since it does not rely on predefined hexahedron triangulations.

\subsection{Pattern occurrences in real meshes}

The theoretical problem of the triangulation
of the hexahedron is an important step for
a better understanding of the methods combining elements
of a tetrahedral mesh to build hexahedra.
Indeed, in practice, when combining the element of a tetrahedral mesh,
many of the additional patterns we identified occur.

\begin{table}
  \centering
  \setlength{\tabcolsep}{3pt}
  %\footnotesize
  \caption{
		Number of triangulations patterns per number of tetrahedra counted in the available input data of \cite{pellerin_identifying_2018}.
		}
	\begin{tabular}{lrcccccccccccc}
		\toprule
		Model  & \#vert .& 5 & 6   & 7  & 8 & 9 & 10 & 11 & 12 & 13 & 14 & 15 & Total \\
 		\midrule
             Cube                & 127               &  1  &   5   &  2  &  1 &  0   &   0  &   0  &    0   &    0  &    0 &  0   &  9      \\
             Fusee               & 11,975            &  1  &   5   &  5  &   7  &  13  &  9   &  4   &   0    &   0   &   0  & 0 &  44      \\
             CShaft       & 23,245        &  1  &   5   &  5  &   7  &  13  &  17   &  6   &   0    &   0   &   0  & 0    & 54   \\
             Fusee\_1          & 71,947          &  1  &   5   &  5  &   7  &  7   &  1   &  0   &   0    &   0   &   0  &  0   &   26    \\
             Caliper             & 130,572          &  1  &   5   &  5  &   7  &  7  &  1   &  0   &   0    &   0   &   0  &  0   &   26 \\
             CShaft2 & 140,985    &  1  &   5   &  5  &   7  &  8  &  0   &  1   &   0    &   0   &   0  & 0    &  27 \\
             Fusee\_2            & 161,888        &  1  &   5   &  5  &    7 &  7  &  1   &  0   &   0    &   0   &   0  &  0   &  26 \\
             FT47\_b    & 221,780       &  1  &   5   &  5  &   7  &  8  &  2   &  0   &   0    &   0   &   0  & 0    &  28 \\
             FT47               & 370,401        & 1   &   5   &  5 &    7 &  7   &  2  &  0    &   0   &   0  &   0    & 0  &   27  \\
             Fusee\_3            & 501,021        &  1  &   5   &  5  &   7  &  8  &  4   &  0   &   0    &   0   &   0  &  0  &  30  \\
             Los1               & 583,561         &  1  &   5   &  5  &   7  &  7  &  2   &  0   &   0    &   0   &   0  & 0  &   27 \\
             Knuckle             & 3,058,481      &  1  &   5   &  5  &  7  &  8  &  2   &  0   &   0    &   0   &   0  &  0  &   28 \\
	\bottomrule
	\end{tabular}
  \label{tab:number_patterns_imr_data}%
\end{table}

\begin{table}
  \centering
  \setlength{\tabcolsep}{3.5pt}
 % \footnotesize
 \caption{
	Number of triangulations patterns per number of tetrahedra counted in the Delaunay triangulations of random point sets.
   }
\begin{tabular}{rcccccccccccc}
	\toprule
	\#vertices   & 5 & 6   & 7  & 8 & 9 & 10 & 11 & 12 & 13 & 14 & 15 & Total \\
	\midrule
            3,000                  &  1  &   5   &  2  &   7  &   13   &  16   &   4  &    0   &    0  &    0 &  0   &  51      \\
            10,000                &  1  &   5   &  5  &   7  &   13   &  19    &  10   &   2    &   0   &   0  & 0 &  62      \\
            20,000                &  1  &   5   &  5  &   7  &   13   &  19    &  15   &   2    &   0   &   0  & 0    & 67   \\
            100,000              &  1  &   5   &  5  &   7  &   13   &  20   &  24   &   5    &   0   &   0  &  0   &   80   \\
            500,000              &  1  &   5   &  5  &   7  &   13  &   20   &  28   &   12    &   1   &   0  &  0   &   92 \\
            1,000,000           &  1  &   5   &  5  &   7  &   13  &   20   &  30   &   14    &   0   &   0  & 0    &  95 \\
            2,000,000           &  1  &   5   &  5  &    7 &   13  &   20   &  30   &   15    &   4   &   1  &  0   &  101 \\
            5,000,000           &  1  &   5   &  5  &   7  &   13  &   20   &  30   &   16    &   4   &   1  & 0    &  102 \\
            10,000,000         & 1   &   5   &  5  &   7 &   13   &   20  &  31    &   16   &   6  &   1    & 0  &   105  \\
\bottomrule
\end{tabular}
 \label{tab:number_patterns_random}%
 \end{table}%

To identify combinations of tetrahedra into hexahedra in a input tetrahedral mesh,
we used the algorithm proposed by \citep{pellerin_identifying_2018}.
This algorithm provably computes all the possible combinations of eight vertices whose connectivity matches the one of a hexahedron.
It further guarantees that all the identified hexahedra are valid for finite element computations
using the exact test of \citep{johnen_geometrical_2013}.
We used the available implementation of \citep{pellerin_identifying_2018}
to compute all valid hexahedra for the input tetrahedral meshes provided
with that paper (Table~\ref{tab:number_patterns_imr_data}) as well as for the Delaunay triangulations of random point sets (Table~\ref{tab:number_patterns_random}).
The decomposition graphs of all identified hexahedra  are compared to the 171 possible triangulations and classified.

There is a clear link between the quality of the point set for hexahedral mesh generation
and the number of patterns of combinations of tetrahedra into hexahedra.
In random point sets, all existing realizable patterns up to 9 tetrahedra are found with as few
as 3,000 randomly distributed vertices.
In point sets generated for hexahedral meshing with the placement strategy
described in \cite{baudouin_frontal_2014},
the number of patterns is smaller and we note that the number of combination patterns is relatively independent of
both the model and  the number of vertices.
We further observed that occurrences of triangulations with more than 11 tetrahedra are rare.

\subsection{Future Work}

Using our results, we would be able to decide on the existence of a triangulation
of eight vertices whose boundary matches a hexahedron. We could also produce all existing triangulations.
Since the principle of our demonstration is general, it could be extended to enumerate the
triangulations of any polyhedron with quadrilateral and triangular faces
and could be used to characterize non-triangulable polyhedra, a major difficulty
in many geometric and combinatorial problems
about which little is known theoretically \citep{rambau_generalization_2003}.
We expect the theoretical advance of this paper to be an important step toward a more complete study
of the geometrical properties of hexahedral cells
and will help develop robust hexahedral mesh generation algorithms.

\begin{acks}
The authors thank the anonymous referees for their valuable comments
and extremely helpful suggestions.

This project has received funding from the European Research Council (ERC) under the European Union's Horizon 2020
research and innovation programme (grant agreement ERC-2015-AdG-694020).

\end{acks}

%\bibliographystyle{ACM-Reference-Format}
%\bibliography{patterns}

%%% -*-BibTeX-*-
%%% Do NOT edit. File created by BibTeX with style
%%% ACM-Reference-Format-Journals [18-Jan-2012].

%\bibliographystyle{ACM-Reference-Format}
%\bibliography{patterns}

\end{document}